%% file: GardestigPhillips.tex
\documentclass[12pt,a4paper]{conference}

\usepackage{fancyhdr}
\usepackage{graphicx,amsmath,amssymb,cite}
\usepackage{multind}
\makeindex{author} \makeindex{subject}

\input MENU07macros.tex

\begin{document}
\Chapter{Precision extraction of \boldmath$a_{nn}$ from 
\boldmath$\pi^-d\to nn\gamma$ using chiral perturbation theory}
        {Precision extraction of $a_{nn}$ from $\pi^-d\to nn\gamma$ using 
$\chi$PT}
	{A. G{\aa}rdestig \it{et al.}}
\vspace{-6 cm}\includegraphics[width=6 cm]{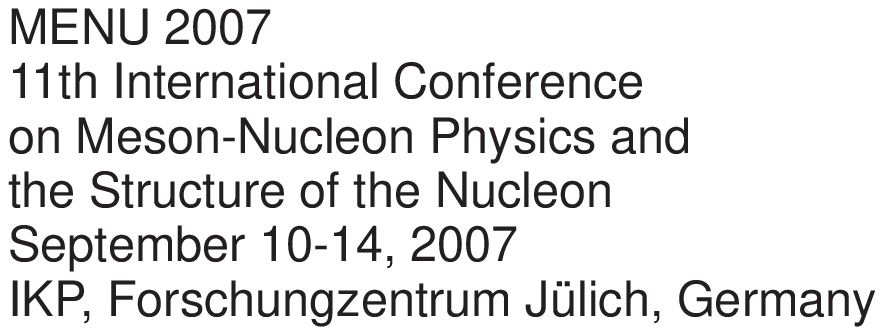}
\vspace{4 cm}
\addcontentsline{toc}{chapter}{{\it A. G{\aa}rdestig}} \label{authorStart}
\begin{raggedright}

{\it A. G{\aa}rdestig}
\index{author}{G{\aa}rdestig, A.}\\
Department of Physics and Astronomy\\
University of South Carolina\\
Columbia, SC 29208\\
U. S. A.
\bigskip\bigskip

{\it D. R. Phillips}
\index{author}{Phillips, D.R.}\\
Department of Physics and Astronomy\\
Ohio University\\
Athens, OH 45701\\
U. S. A.
\bigskip\bigskip

\end{raggedright}

\begin{center}
\textbf{Abstract}
\end{center}
The \index{subject}{neutron-neutron scattering length}
neutron-neutron scattering length $a_{nn}$ provides a sensitive
probe of \index{subject}{charge-symmetry breaking}charge-symmetry breaking 
in the strong interaction. 
Here we summarize our recent efforts to use 
\index{subject}{chiral perturbation theory}
chiral perturbation theory in order to systematically relate $a_{nn}$ to 
the shape of the neutron spectrum in the reaction 
\index{subject}{radiative pion capture}
$\pi^- d \rightarrow nn \gamma$.
In particular we show how the chiral symmetry of QCD relates this process 
to low-energy electroweak reactions such as $pp \rightarrow d e^+ \nu_e$. 
This allows us to reduce the uncertainty in the extracted $a_{nn}$
(mainly due to short-distance physics in the two-nucleon system)
by a factor of more than three, to $<0.05$~fm. 
We also report first results on the impact that two-nucleon mechanisms of 
chiral order $P^4$ have on the $\pi^- d \rightarrow nn \gamma$ 
neutron spectrum.

\section{Introduction}
Quantum chromodynamics (QCD) is almost symmetric under the interchange of
the up and down quarks.  
This is called charge symmetry, and is due to the fact that the mass 
difference $m_d - m_u$ is much smaller than the QCD mass scale 
$\Lambda\sim1$~GeV.
This symmetry, which is a subgroup of isospin symmetry $SU(2)_V$, is
well-respected in strong interactions at low energies, but is softly
broken by quark-mass differences, and also by electromagnetic effects. 
The relevant dimensionless parameters governing
charge-symmetry breaking are therefore $\frac{m_d-m_u}{\Lambda}$
and $\frac{\alpha_{em}}{\pi}$, both of which are less than 1\%.
Although it is indeed generally of this small magnitude,
charge-symmetry breaking (CSB) has many experimentally verifiable effects 
in hadronic and nuclear physics, such as the neutron-proton mass difference,
rho-omega mixing, the binding-energy difference of mirror nuclei
(e.g., $^3$He and $^3$H), the recently measured forward-backward
symmetry for $np\to d\pi^0$~\cite{Allena}, and the $dd\to\alpha\pi^0$
reaction~\cite{IUCFCSB}.  
For comprehensive reviews of charge symmetry
and its breaking see Ref.~\cite{MNS}.

The difference between the strong-interaction parts of the $nn$ and $pp$ 
scattering lengths ($a^{\rm str}_{pp}-a^{\rm str}_{nn}$) is
particularly sensitive to CSB.
The scattering length parameterizes the zero-energy $NN$ scattering phase 
shift, $\delta(p)$, via:
\begin{equation}
  a \equiv -\lim_{p \rightarrow 0} \frac{\delta(p)}{p},
\end{equation}
where $p$ is the $NN$ relative momentum.
Hence the (strong) $nn$ and $pp$ scattering lengths would be equal in the 
limit of exact charge symmetry. 
Their difference is an important quantity for two, somewhat related reasons.  
Firstly, the nucleon-nucleon scattering lengths are unnaturally large compared 
to the pion Compton wavelength. 
This is indicative of fine tuning in the $NN$ potential
and in consequence the CSB piece of the $NN$ potential has an impact
on the scattering lengths that is greatly enhanced~\cite{MNS}:
\begin{equation}
\frac{a^{\rm str}_{pp} - a^{\rm str}_{nn}}{a}=(10-15)\frac{\Delta
  V_{\rm CSB}}{V_{NN}}.
\end{equation}  
Measurements of $a_{pp}-a_{nn}$ therefore provide significant constraints on 
CSB terms in modern phenomenological $NN$ potentials, e.g., AV18~\cite{AV18}. 
Secondly, when potentials fit to the currently accepted values
$a_{nn}=-18.59\pm0.4$~fm and $a_{pp}=-17.3\pm0.4$~fm~\cite{NNreview}
are used to make predictions for binding energies of mirror nuclei, 
they very accurately reproduce the experimental binding-energy difference 
of, e.g., the aforementioned ${}^3$H and ${}^3$He~\cite{GFMC}.

Both $a_{nn}$ and $a_{pp}$ must have electromagnetic corrections
applied to them in order to extract the strong-interaction part.
This correction is huge for the $pp$ case, but is under good theoretical
control.
In the $nn$ case the electromagnetic correction is due to a magnetic-moment 
interaction and is $\approx -0.3$~fm.

But on the $nn$ side there is an experimental 
difficulty in obtaining dense enough free nucleon targets.
There have been some attempts at doing direct $nn$ measurements, the most
recent one being pursued at the pulsed reactor YAGUAR~\cite{yaguar}.
However, the more promising approaches so far have been based on indirect 
measurements, where final-state neutrons are detected in regions of phase
space where they have low relative energy and hence observables are
sensitive to the $nn$ scattering length.

Unfortunately, the two most recent measurements employing the $nd\to nnp$ 
reaction for this purpose extract very different $a_{nn}$ values.
Thus, a Bonn group reported $a_{nn}=-16.1\pm0.4$~fm\cite{Huhn}, while a group
based at TUNL claimed $a_{nn}=-18.7\pm0.7$~fm~\cite{TUNL}, a 4$\sigma$ 
disagreement.

However, experiments at different facilities based on the
alternative process $\pi^-d\to nn\gamma$~\cite{Gabioudetal,LAMPF},
have yielded consistent values for many years. 
Thus these results dominate the ``accepted'' value of $a_{nn}$ quoted above.  
The scattering length is extracted by fitting the shape of the spectrum 
of neutrons emitted from the decay of the pionic deuterium atom.
The theoretical uncertainty in the $a_{nn}$ extracted using
extant calculations~\cite{GGS,deTeramond} is $\approx \pm0.3$~fm, 
dominated by the uncertainties in the $nn$ wave function at short distances.

In the present work we revisit these calculations for radiative pion
capture on deuterium and take advantage of the modern development of
effective field theory (EFT), in particular chiral perturbation theory
($\chi$PT).  
By using an EFT we have consistency between the wave
functions and production/capture amplitudes, a recipe to estimate the
theoretical error, and we can make systematic improvements when
necessary.  Also, in the case of $\chi$PT, we gain a close connection
to the underlying theory QCD through QCD's chiral symmetry.  In the
next section we describe the key elements of our $\chi$PT 
calculation of $\pi^- d \rightarrow nn \gamma$, and in 
Section~\ref{sec-results} we present the results
already obtained in recent publications~\cite{GP1,GP2,AG1}, and also
provide a first report on substantial improvements of these
calculations.

\section{Anatomy of the Calculation}

EFTs circumvent the problem of the large QCD coupling constant at low energies.
Instead one expands amplitudes in the ratio $P/\Lambda$, where $P\sim m_\pi$ 
is a small energy/momentum of the problem and $\Lambda\sim1$~GeV is the scale 
of chiral-symmetry breaking. 
This power counting provides a hierarchy of quantum-mechanical amplitudes 
which allows for an systematic organization of the calculation.  
(We count the electron charge $e$ as one power of $P/\Lambda$.)

In the application of 
chiral perturbation theory to nuclear processes, classes of graphs must be 
resummed in order to generate the nuclear bound states observed in nature. 
The original proposal for such resummation is due to Weinberg~\cite{Weinberg}. 
Applying it to the case at hand we see that the amplitude for 
$\pi^- d \rightarrow nn \gamma$ should be calculated as:
\begin{equation}
  {\cal A}=\langle {\bf p}|\hat{O}|\psi_d \rangle + 
  \langle {\bf p}|T_{nn} G_0 \hat{O} |\psi_d \rangle,
\end{equation}
where $|\psi_d \rangle$ is the deuteron wave function (which is
dominated by modes with momenta where $\chi$PT is applicable), 
$|{\bf  p}\rangle$ is a plane wave with the observed relative momentum of
the two final-state neutrons, ${\bf p}$, $T_{nn}$ is the $nn$
rescattering amplitude, and $G_0$ the free $nn$ Green function.

Meanwhile $\hat{O}$ is the operator (technically the two-particle
irreducible kernel) governing the transition $\pi^- np \rightarrow nn
\gamma$. Weinberg proposed that $\hat{O}$ has a well-behaved chiral
expansion and so can be calculated in $\chi$PT. (For a summary of the
successful application of this idea to electromagnetic processes see
Ref.~\cite{Ph07}.)  $\hat{O}$ has one- and two-body pieces, with the
one-body part in this case beginning at ${\cal O}(P)$ with the
Kroll-Ruderman term for $\pi^- p \rightarrow n \gamma$. 
Two-body pieces enter at ${\cal O}(P^3)$. 
In this work we report on calculations obtained from a partial ${\cal O}(P^4)$
(next-to-next-to-next-leading order = N3LO) calculation of $\hat{O}$.
Our calculation includes all mechanisms at ${\cal O}(P^3)$ (N2LO), but 
only the dominant ${\cal O}(P^4)$ two-body pieces of $\hat{O}$.

\subsection{Chirally inspired wave functions}

In order to reach the desired accuracy in the calculation of $\cal{A}$, the
$NN$ wave functions have to be calculated to an order that is consistent with
that to which $\hat{O}$ is obtained.
Here this means that they must be computed up to ${\cal O}(P^3)$ and 
thus include the leading- and sub-leading two-pion-exchange 
corrections to the chiral $NN$ potential~\cite{chiTPE}.  
The necessary deuteron and $nn$ scattering
wave functions are derived starting from the asymptotic states, given
by the asymptotic normalization $A_S$ and $D/S$ ratio for the deuteron
and the effective-range expansion for $nn$ scattering.  
These are integrated in from $r=\infty$ using the Schr\"odinger equation with
the chiral one- and two-pion exchange potentials.  Eventually, we
reach a region, at $r=1$--$2$ fm, where the chiral expansion
for the $NN$ potential breaks down.  We take the simple approach of
introducing a cutoff $R$ in this range and assume that the
potential for $r<R$ is given by a square well whose depth we adjust
to enforce continuity of the wave function at $r=R$.  
This parameterizes and regularizes our ignorance of the short-distance $NN$
physics.  
It is then important to ensure that the result is
independent of the cutoff $R$ to the order we are working, i.e., that 
the renormalization-group criteria are fulfilled.

\subsection{One-body amplitudes to NNLO}
The chiral one-body amplitudes have been calculated by 
Fearing {\it et al.}~\cite{Fearing} up to $\mathcal{O}(P^3)$, fitting the 
available $\gamma p\rightarrow \pi^+ n$ and $\pi^- p \rightarrow \gamma n$ data
via the  adjustment of $\chi$PT low-energy constants (LECs).  A preliminary
estimate of the size of the N3LO one-body amplitude indicates that it
has negligible influence on $\pi^-d\to nn\gamma$ and hence
we do not discuss it further here~\cite{GP3}.

\subsection{Two-body amplitudes to N3LO}
\begin{figure}[ht]
\begin{center}
\includegraphics[width=5.4in]{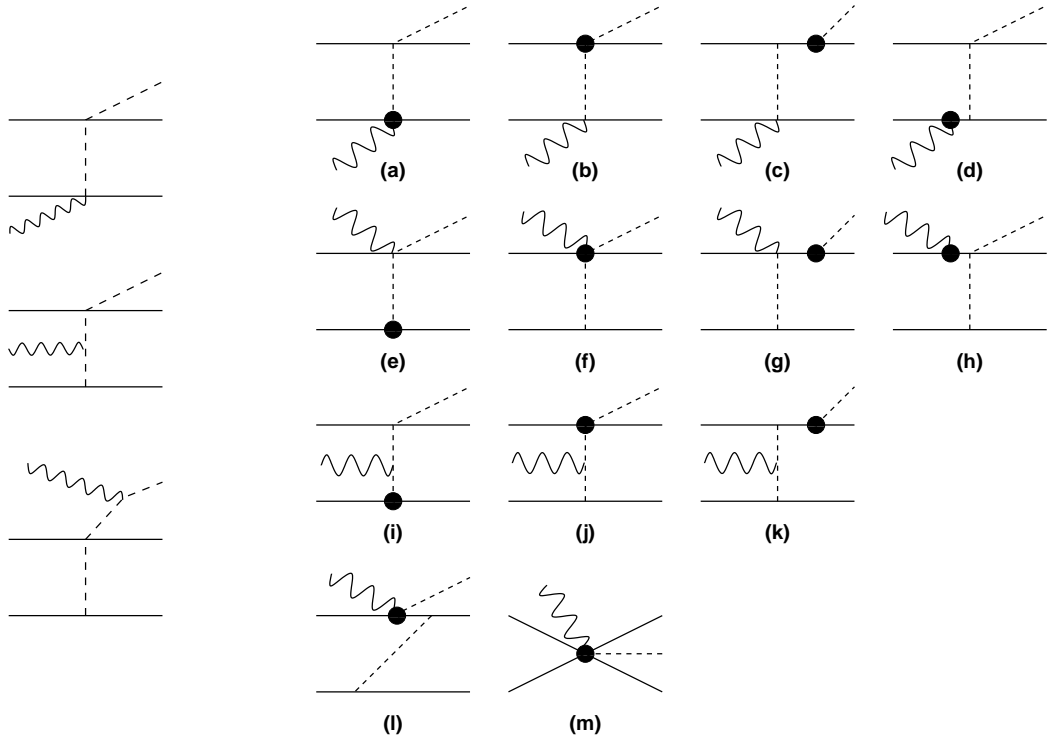}
\caption{The two-body diagrams relevant for $\pi^-d\to nn\gamma$.
Left: ${\cal O}(P^3)$.
Right: $\mathcal{O}(P^4)$.
Only one representative vertex ordering is given for each type of diagram.
The black discs indicates insertions from $\mathcal{L}^{(2)}_{\pi N+\pi NN}$.}
\label{fig:OQ34tb}
\end{center}
\end{figure}
At $\mathcal{O}(P^3)$ (N2LO), there are the three diagrams given to the left 
in Fig.~\ref{fig:OQ34tb}.  The first is believed to be larger than the
second since the pion mass disappears from its single propagator and the
pion can go on-shell, while the second has, in addition, also one off-shell 
propagator. 
The third diagram vanishes in Coulomb gauge.
At N3LO [$\mathcal{O}(P^4)$], a slew of diagrams appear, given to the right in 
Fig.~\ref{fig:OQ34tb} and discussed in detail in~\cite{AG1}. 
The overall result for the neutron time-of-flight spectrum when all these
graphs except for $(l)$ are included is depicted in Fig.~\ref{fig:RdepN23LO}.
Diagram $(l)$ is related to the orthonormalization of the wave functions.
Since this is suppressed by $1/M$, we expect that effects due to $c_2$--$c_4$, 
which appear in the two-body currents computed in Ref.~\cite{AG1} that are 
included in our calculation, will be substantially larger. 
A complete calculation of the ${\cal O}(P^4)$ correction, including  
orthonormalization, is under way~\cite{GP3}.

\subsection{Constraining unknown short-distance physics}

Fig.~\ref{fig:RdepN23LO} shows that the neutron spectrum
calculated at ${\cal O}(P^4)$ with different values of the regulator radius
$R$ and a fixed value of the short-distance coefficient in diagram
$(m)$ is significantly different in the final-state-interaction (FSI) region. 
Since $a_{nn}$ is extracted by fitting the shape of the spectrum in the
FSI region~\cite{LAMPF} this sensitivity to unconstrained physics of the $NN$ 
system seems to limit the accuracy with which $a_{nn}$ can be obtained from 
$\pi^- d \rightarrow nn \gamma$. 
We now show how to remedy this problem.

The LO contribution to the matrix element in this region is given by 
\begin{equation}
  {\cal M}_{FSI} \equiv C \int_0^\infty dr \, u_{nn}(r;p) j_0 
  \left(\frac{k r}{2} \right) u_d(r),
\label{eq:MFSI}
\end{equation}
where $k$ is the momentum of the outgoing photon, $C$ is a constant
that is fixed by $e$, $g_A$, $f_\pi$, etc., $j_0$ is the spherical
Bessel function of zeroth order, and $u_d$ [$u_{nn}(r;p)$] is the radial S-wave
wave function of the deuteron (${}^1S_0$) state.
But the short-distance part of this matrix element is 
the same as that of the $pp$ fusion matrix element
\begin{equation}
{\cal M}_{GT} \equiv \int_0^\infty dr \, u_{pp}(r) u_d(r).
\end{equation}%
\begin{figure}[ht]
\begin{center}
\includegraphics*[width=4in]{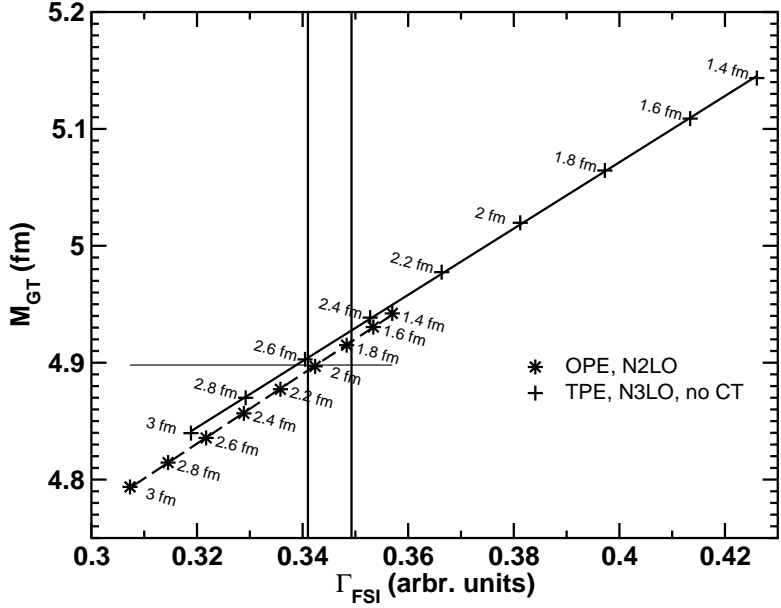}
\caption{Gamow-Teller matrix element plotted against $\pi^-d\to nn\gamma$ FSI 
  peak height for various values of $R$ and at different orders and 
  wave functions.
  The points correspond to the values of $R$ (in fm) indicated.
  The straight lines are linear fits to the points.
  The vertical lines show the range in $\Gamma_{\rm FSI}$ after 
  renormalization, given the $M_{\rm GT}$ value indicated by the
  horizontal line.}
\label{fig:GPline}
\end{center}
\end{figure}%
This connection is shown empirically in Fig.~\ref{fig:GPline}, revealing a 
linear relationship between the Gamow-Teller matrix element 
${\cal M}_{\rm GT}$, and the FSI peak height in $\pi^-d\to nn\gamma$, 
$\Gamma_{\rm FSI}\sim|{\cal M}_{\rm FSI}|^2$. 

This can be understood from the structure of the
chiral Lagrangian.  It contains both one-nucleon and two-nucleon terms
linear in the axial field $u_\mu$:
\begin{eqnarray}
\mathcal{L} & = & N^\dagger(iv\cdot D+g_{\rm A}S\cdot u)N \nonumber \\
& - & 2d_1N^\dagger S\cdot uNN^\dagger N
  +2d_2\epsilon^{abc}\epsilon_{\kappa\lambda\mu\nu}v^\kappa
  u^{\lambda,a}N^\dagger S^\mu\tau^b NN^\dagger S^\nu\tau^c N\ldots,
\end{eqnarray}
where
\begin{equation}
f_\pi u_\mu = -\tau^a \partial_\mu \pi^a-\epsilon^{3ba}V_\mu \pi^b \tau^a
+f_\pi A_\mu + \mathcal{O}(\pi^3),
\end{equation}
$V_\mu$ $(A_\mu)$ is an external vector (axial) field, and the 
$d_i=\mathcal{O}(\frac{1}{Mf_\pi^2})$ are (a priori unknown) LECs. 
Since $u_\mu$ contains the pion pseudovector coupling, as well as a pion-photon
coupling and the axial field $A_\mu$, this is the chiral explanation behind 
the well-known \index{subject}{Goldberger-Treiman}Goldberger-Treiman (GT) 
relation and \index{subject}{Kroll-Ruderman}Kroll-Ruderman (KR) terms. 
In the two-nucleon sector, the same features of $u_\mu$ imply a connection 
between pion $p$-wave production, pion photoproduction on the $NN$ system, and 
axial currents---two-body analogs of the GT and KR.
The connection between pion production and electroweak processes is currently
being investigated by Nakamura~\cite{SXN}.

For \index{subject}{Gamow-Teller}Gamow-Teller ($^1S_0\leftrightarrow{}^3S_1$)
transitions, the LECs only appear in the combination
\begin{equation}
  \hat{d} \equiv \hat{d}_1+2\hat{d}_2+\frac{\hat{c}_3}{3}+\frac{2\hat{c}_4}{3}
+\frac{1}{6},
\end{equation}
where $g_{\rm A}\hat{d}_i\equiv Mf_\pi^2d_i$ and $\hat{c}_i\equiv
Mc_i$~\cite{Parkhep}.  This LEC also appears in $p$-wave pion
production in $NN$ collisions, tritium $\beta$ decay, $pp$ fusion,
$\nu d$ scattering, $\mu^-d\to nn\nu_\mu$, and the hep reaction.  In
addition, if the emitted pion couples to a third nucleon, this same
operator and coefficient enters the leading 
\index{subject}{three-nucleon force, chiral} chiral three-nucleon force.  
These and other implications are
discussed further in Refs.~\cite{GP2,AG1}. The key point in the
context of this work is that chiral symmetry and gauge invariance
together explain the linear correlation between $pp$ fusion and the
FSI peak height that is evident in Fig.~\ref{fig:GPline}.

Now, the \index{subject}{solar fusion} solar $pp$ fusion process has
recently been calculated very accurately by constraining its unknown 
short-distance physics from precise calculations of tritium beta 
decay~\cite{Parkhep}. 
If we adjust the LEC that appears in diagram $(m)$  to reproduce 
this rate for $pp\to de^+\nu_e$ we obtain a very precise prediction for 
the FSI peak height in $\pi^- d \rightarrow nn \gamma$, as shown in 
Fig.~\ref{fig:GPline}.

\section{Results}
\label{sec-results}
\begin{figure}[ht]
\begin{center}
\includegraphics*[width=4.2in]{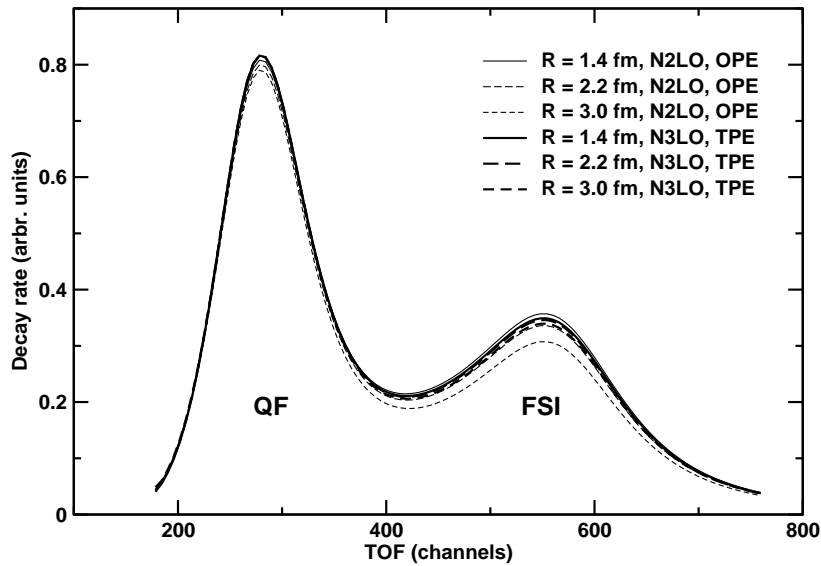}
\caption{The neutron time-of-flight spectrum for $\pi^-d\to nn\gamma$ at
different cutoffs $R$ and orders as indicated.
The thin lines are for the N3LO calculation with wave functions calculated with
the chiral one-pion exchange potential, while the thick lines include N3LO
two-body currents and the chiral two-pion exchange potential as well.
The latter coincide at the quasi-free (QF) peak and show a much reduced spread 
in the FSI peak.}
\label{fig:RdepN23LO}
\end{center}
\end{figure}
The result of this renormalization can be seen in
Fig.~\ref{fig:RdepN23LO}. 
Clearly the N3LO contribution reduces the cutoff dependence considerably 
compared to N2LO. The theoretical uncertainty due to unknown
short-distance physics in the $NN$ system is now negligible in the
FSI region. A detailed analysis of the other theoretical uncertainties
(see Ref.~\cite{GP1}) reveals that 
the total theoretical error in the extracted $a_{nn}$ at N3LO is $\pm0.3$~fm 
when the entire spectrum is fitted and $\pm0.05$~fm if only the FSI peak is 
fitted.

\section{Conclusions}
Chiral perturbation theory relates the unknown
short-distance physics of various electroweak two-body observables to
pion $p$-wave production and pion photoproduction on two nucleons.
(We can also constrain a piece of the chiral three-nucleon force from 
electroweak two-body observables.)  
This connection makes it possible to calculate $\pi^-d\to nn\gamma$
to high precision, leading to a small theoretical error for
the extraction of $a_{nn}$: $\Delta a_{nn}^{\rm theory}=\pm0.05$~fm.
This reduces that error by at least a factor of three compared to
previous calculations.

A future publication~\cite{GP3} will contain a full description of the
amplitudes and wave functions employed in our N3LO calculation. In
that work we will also investigate the influence of higher-order
electromagnetic corrections in the $pp$ wave functions used for 
$pp\to de^+\nu_e$ and whether we are justified in neglecting the N3LO 
one-body contribution.
We also provide a full accounting of the $1/M$ corrections to the
two-body operators that are mandated by the unitary transformations
used to obtain a Hermitian $NN$ potential $V$. 

In addition we are investigating the possibility to constrain
$\hat{d}$ directly from a two-body observable by calculating the
$\mu^-d\to nn\nu_\mu$ capture rate in the same framework~\cite{mud}.
This reaction is soon to be measured at the Paul Scherrer Institute to
1\% precision~\cite{Kammel}.  It would also be interesting to revisit
the neutrino-deuteron breakup reactions that are important for the SNO
detector.  Another possible direction would be to complete the circle
by calculating tritium beta decay using chiral three-nucleon wave
functions with the $r$-space regularization we have used in the $NN$
sector.

\section*{Acknowledgments}
This work was supported by the Institute for Nuclear and Particle Physics at
Ohio University, by DOE grant DE-FG02-93ER40756, and by 
NSF grant PHY-0457014.
\appendix

\end{document}

%% file: MENU07macros.tex
\newcommand{\beq}{\begin{equation}}
\newcommand{\eeq}[1]{\label{#1}\end{equation}}
\newcommand{\eeqn}{\end{equation}}

\newcommand{\beqa}{\begin{eqnarray}}
\newcommand{\eeqa}[1]{\label{#1}\end{eqnarray}}
\newcommand{\eeqan}{\end{eqnarray}}

\let\bar=\overbar

\newcommand{\Dslash}{\not{\hbox{\kern-4pt $D$}}}
\newcommand{\dslash}{\not{\hbox{\kern-2pt $\del$}}}

\newcommand{\msb}{{\bar{\ssstyle M \kern -1pt S}}}




%% file: GardestigPhillips.bbl
\begin{thebibliography}{00}

\bibitem{Allena} 
A.~K.~Opper {\it et al.}, 
{\it Phys.\ Rev.\ Lett.}\ {\bf 91}, 212302 (2003).

\bibitem{IUCFCSB} 
E.~J.~Stephenson {\it et al.}, 
{\it Phys.\ Rev.\ Lett.}\ {\bf 91}, 142302 (2003).

\bibitem{MNS}
G.~A.~Miller, B.~M.~K.~Nefkens, and I.~\v{S}laus, 
{\it Phys.\ Rep.}\ {\bf 194}, 1 (1990);
G. A. Miller and W. van Oers,
arXiv:nucl-th/9409013;
G. A. Miller, A. Opper, and E. J. Stephenson, 
{\it Ann.\ Rev.\ Nucl.\ Part.\ Sci.}\ {\bf 56}, 293 (2006).

\bibitem{AV18}
R.~B. Wiringa, V.~G.~J. Stoks, and R.~Schiavilla,
{\it Phys.\ Rev.\ C} {\bf 51}, 38 (1995).

\bibitem{NNreview}
R.~Machleidt and I.~Slaus,
{\it J. Phys.\ G} {\bf 27}, R69 (2001).

\bibitem{GFMC}
S.~C.~Pieper and R.~B.~Wiringa,
{\it Annu.\ Rev.\ Nucl.\ Part.\ Sci.}\ {\bf 51}, 53 (2001).

\bibitem{yaguar}
W.~I.~Furman {\it et al.},
{\it J. Phys.\ G: Nucl.\ Part.\ Phys.}\ {\bf 28}, 2627 (2002).

\bibitem{Huhn}
V.~ Huhn {\it et al.},
{\it Phys.\ Rev.\ Lett.}\ {\bf 85}, 1190 (2000).

\bibitem{TUNL}
D.~E.~Gonz\'alez~Trotter {\it et al.},
{\it Phys.\ Rev.\ Lett.}\ {\bf 83}, 3788 (1999).
{\it Phys.\ Rev.}\ C {\bf 73}, 034001 (2006).

\bibitem{Gabioudetal}
B.~Gabioud {\it et al.},
{\it Phys.\ Rev.\ Lett.}\ {\bf 42}, 1508 (1979);
{\it Phys.\ Lett.}\ {\bf 103B}, 9 (1981);
{\it Nucl.\ Phys.}\ {\bf A420}, 496 (1984);
O.~Schori {\it et al.},
{\it Phys.\ Rev.}\ C {\bf 35}, 2252 (1987).

\bibitem{LAMPF}
C.~R.~Howell {\it et al.},
{\it Phys.\ Lett.\ B} {\bf 444}, 252 (1998).

\bibitem{GGS}
W.~R.~Gibbs, B.~F.~Gibson, and G.~J.~Stephenson, Jr.,
{\it Phys\ Rev.}\ C {\bf 11}, 90 (1975); 
{\bf 16}, 322 (1977); 
{\bf 16}, 327 (1977).

\bibitem{deTeramond}
G.~F.~de~T\'eramond,
{\it Phys.\ Rev.}\ C {\bf 16}, 1976 (1977);
G.~F.~de~T\'eramond, J.~P\'aez, and C.~W.~Soto~Vargas,
{\it ibid.} {\bf 21}, 2542 (1980);
G.~F.~de~T\'eramond and B.~Gabioud,
{\it ibid.} {\bf 36}, 691 (1987).

\bibitem{GP1}
A.~G{\aa}rdestig and D.~R.~Phillips,
{\it Phys.\ Rev.\ C} {\bf 73}, 014002 (2006).

\bibitem{GP2}
A.~G{\aa}rdestig and D.~R.~Phillips,
{\it Phys.\ Rev.\ Lett.}\ {\bf 96}, 232301 (2006).

\bibitem{AG1}
A.~G{\aa}rdestig,
{\it Phys.\ Rev.\ C} {\bf 74}, 017001 (2006).

\bibitem{Weinberg}
  S.~Weinberg,
  {\it Nucl.\ Phys.\ B} {\bf 363}, 3 (1991);
  {\it Phys.\ Lett.\ B} {\bf 251}, 288 (1990).

\bibitem{Ph07}
D.~R.~Phillips, these proceedings.

\bibitem{chiTPE}
C.~Ordon\'ez, L.~Ray, and U. van Kolck, 
{\it Phys.\ Rev.}\ C {\bf 53}, 2086 (1996);
N.~Kaiser, R.~Brockmann, and W.~Weise, 
{\it Nucl.\ Phys.}\ {\bf A625}, 758 (1997);
E.~Epelbaum, W.~Gl\"ockle, and U.-G.~Mei\ss ner, 
{\it Nucl.\ Phys.}\ {\bf A671}, 295 (1999);
M.~C.~M.~Rentmeester, {\it et al.},
{\it Phys.\ Rev.\ Lett.}\ {\bf 82}, 4992 (1999).

\bibitem{Fearing}
H.~W.~Fearing {\it et al.},
{\it Phys.\ Rev.}\ C {\bf 62}, 054006 (2000).

\bibitem{GP3}
A. G{\aa}rdestig and D. R. Phillips,
(in preparation).

\bibitem{SXN}
S. X. Nakamura, arXiv:0709.1239 [nucl-th] and these proceedings.

\bibitem{Parkhep}
T.-S.~Park {\it et al.},
{\it Phys.\ Rev.}\ C {\bf 67}, 055206 (2003).

\bibitem{mud}
A. G{\aa}rdestig, T.-S. Park, K. Kubodera, and F. Myhrer,
(in preparation).

\bibitem{Kammel}
P. Kammel, talk at  {\it International Conference on Muon 
Catalyzed Fusion and Related Topics}, Dubna, Russia (2007).

\end{thebibliography}
